\documentclass[10pt,a4paper]{article}
\pdfoutput=1
\usepackage[utf8]{inputenc}
\usepackage{amsmath}
\usepackage{amsfonts}
\usepackage{amssymb}

\usepackage[a4paper]{geometry}
\usepackage{graphicx}
\usepackage[textwidth=8em,textsize=small]{todonotes}
\usepackage{amsmath}
\usepackage{natbib}

\title{Projecting UK Mortality using Bayesian Generalised Additive Models}
\author{Jason Hilton, Erengul Dodd, Jon Forster, Peter W. F. Smith\\
Centre for Population Change,\\
University of Southampton,\\
Southampton,\\
United Kingdom,\\
SO17 1BJ \\
J.D.Hilton@soton.ac.uk
}

\begin{document}
\maketitle
\begin{abstract}
Forecasts of mortality provide vital information about future populations, with implications for  pension and health-care policy as well as for decisions made by private companies about life insurance and annuity pricing. Stochastic mortality forecasts allow the uncertainty in mortality predictions to be taken into consideration when making policy decisions and setting product prices. Longer lifespans imply that forecasts of mortality at ages 90 and above will become more important in such calculations.

This paper presents a Bayesian approach to the forecasting of mortality that jointly estimates a Generalised Additive Model (GAM) for mortality for the majority of the age-range and a parametric model for older ages where the data are sparser. The GAM allows smooth components to be estimated for age, cohort and age-specific improvement rates, together with a non-smoothed period effect.  Forecasts for the United Kingdom are produced using data from the Human Mortality Database spanning the period 1961-2013. A metric that approximates predictive accuracy under Leave-One-Out cross-validation is used to estimate weights for the `stacking' of forecasts with different points of transition between the GAM and parametric elements.

Mortality for males and females are estimated separately at first, but a joint model allows the asymptotic limit of mortality at old ages to be shared between sexes, and furthermore provides for forecasts accounting for correlations in period innovations. The joint and single sex model forecasts estimated using data from 1961-2003 are compared against observed data from 2004-2013 to facilitate model assessment.
\end{abstract}

\section*{Acknowledgments}\label{acknowledgments}

This work was supported by the ESRC Centre for Population Change - phase II (grant ES/K007394/1), and a research contract (“Review of Mortality Projections”) between the Office of National Statistics and the University of Southampton. The use of the IRIDIS High Performance Computing Facility, and associated support services at the University of Southampton, in the completion of this work is also acknowledged. Earlier work on this model was presented at a joint Eurostat/UNECE work session on demographic projections \citep{Forster2016}. All the views presented in this paper are those of the authors only.

\section{Introduction}\label{introduction}

The future level of mortality is of vital interest to policy makers and
private insurers alike, as lower mortality results in greater
expenditure on pension payments and higher social care spending.
Individuals are living longer due to improved mortality conditions and
will reach higher ages in greater number as the post-war baby-boom
cohort ages, and thus forecasts of mortality at the oldest ages are
becoming more important. However, these remain challenging to produce,
as the available mortality data at these ages are sparse and
concentrated in the most recent years. The work of \citet{Dodd2018} in
producing the 17\textsuperscript{th} iteration of the English Life
Tables provided a methodology for mortality estimation that combines
smoothing based on Generalised Additive Models (GAMs) \citep{Wood2006}
at the youngest ages with a parametric model at older ages. This paper
extends this approach to a forecasting context and introduces period and
cohort effects, producing fully probabilistic mortality projections
within a Bayesian framework.

\section{Mortality Forecasting}\label{mortality-forecasting}

\subsection{Mortality Rates}\label{mortality-rates}

The raw materials for stochastic mortality forecasts are data on the
number of deaths \(d_{xt}\) in year \(t\) and age last birthday \(x\),
and matching population counts \(P_{xt}\) derived from census data
adjusted for births, deaths and migration in the intervening period. The
appropriate exposures to risk, needed for the calculation of mortality
rates, can be estimated from these population counts. Most often, the
estimated mid-year population totals \(P_{x(t + 0.5)}\) are used to
directly approximate exposures over the whole year \(R_{xt}\), under the
assumption that births, deaths and migrations occur uniformly throughout
the year.

\begin{figure}[htbp]
\centering
\includegraphics{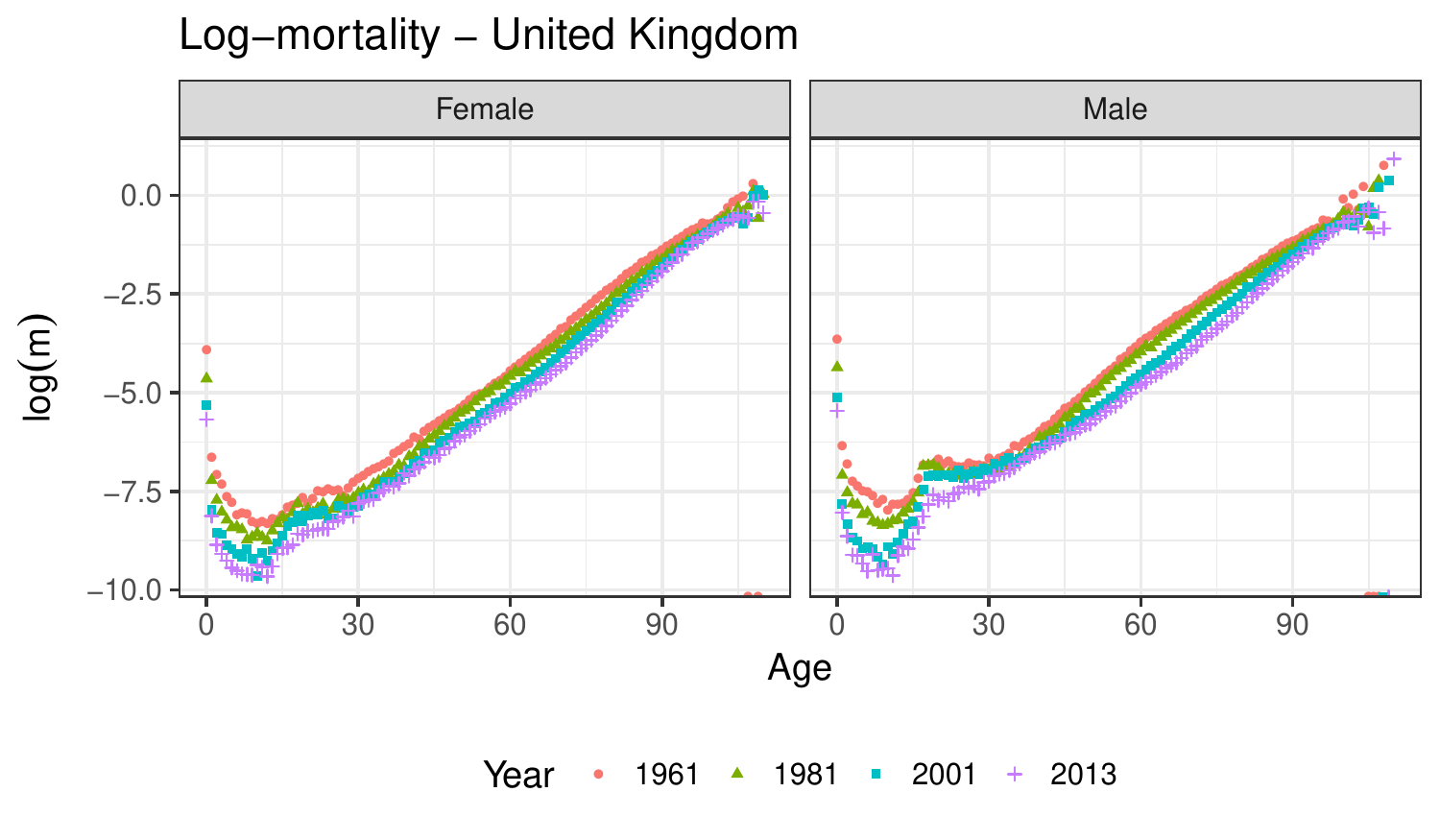}
\caption{Log-Mortality rates for the United Kingdom for selected years,
males and females. Source: Human Mortality
Database\label{fig:empirical_plots}}
\end{figure}

The observed deaths rates \(d_{xt} / R_{xt}\) for the United Kingdom for
the years 1961, 1981, 2001 and 2013 are displayed in Figure
\ref{fig:empirical_plots}, based on data taken from the Human Mortality
Database \citep{hmd}. The Human Mortality Database uses a more
sophisticated method of approximating exposure to risk than that
described above, accounting for the distribution of deaths within single
years of age \citep{Wilmouth2017}. The plotted mortality rates can be
seen to decrease with time, and consistently increase with age beyond
early adulthood, as might be expected. The empirical rates appear
volatile at higher ages where there are fewer survivors and therefore
less data.

The central mortality rate, the quantity which we wish to estimate and
forecast, is defined as \begin{equation}
m_{xt} = \frac{E[d_{xt}]}{R_{xt}}.
\label{eq:cmr}\end{equation}

This is equal to the force of mortality or hazard of death \(\mu(x)\)
within the year and age group under the assumption that the force of
mortality is constant over that interval. \citet{Thatcher1998} and
\citet{Keyfitz2005} provide more detail on the exact relationship
between these quantities.

\subsection{Models of Mortality}\label{models-of-mortality}

A large part of the existing literature on stochastic mortality
modelling has developed from the work of \citet{leecarter1992}. This
approach models the log-mortality rate \(\text{log}(m_{xt})\) using an
age-specific term \(\alpha_{x}\), giving the mean mortality rate for
each age \(x\), and a bi-linear term \(\beta_x\kappa_t\), where the
\(\kappa\) vector describes the overall pace of mortality decline, while
the \(\beta\) coefficients describe how this decline varies by age, so
that

\begin{equation}
\text{log}({m_{xt}}) = \alpha_{x} + \beta_x\kappa_t.
\label{eq:lc}\end{equation}

This reduces the complexity of the forecasting problem, as only the
\(\kappa\) component varies over time. This can be modelled using
standard Box-Jenkins methods (most often a random walk with drift),
which also provide for measures of forecast uncertainty.

The simplicity of the Lee-Carter model has led to a large range of other
adjustments and extensions. \citet{Brouhns2002}, for example, estimate
the parameters through maximisation of a Poisson likelihood for the
observed deaths rather than working with a Gaussian likelihood on the
log-rates, as in the original paper. \citet{rhageenhance}, in contrast,
include multiple bi-linear age-period terms to capture a greater
proportion of the total variation than is possible with a single term.

\citet{Renshaw2006} go further by adding a cohort term
\(\beta^{(2)}_x\gamma_{t-x}\) to allow for differences in mortality by
year of birth. Models that include cohort terms are attractive as in
some countries, and notably in the United Kingdom, cohort effects are
prevalent in the underlying mortality data, possibly reflecting the
different life experiences and lifestyle habits of those born in
different periods \citep{Willets2004, Cairns2009}. Standard
Age-Period-Cohort (APC) models can therefore capture such
characteristics of the data, but given the linear dependence in such
models (in that \(c=t-x\), with \(c\) indexing cohort), identifying
constraints are needed for fitting.

The work of Cairns and collaborators \citep{Cairns2009, Dowd2010}
describes a family of models where mortality is modelled through sums of
terms of the form \(\beta_x\kappa_t\gamma_{t-x}\), where \(\beta_x\)
refers to age effects, \(\kappa_t\) to period effects, and
\(\gamma_{t-x}\) to cohort effects. Any of these elements may be
constant or deterministic in particular models, and so the Lee-Carter
and Age-Period-Cohort models are incorporated as special cases. The
in-sample and forecasting performance of these models are assessed
against a number of criteria in \citet{Cairns2009}. A notable finding
was the lack of robustness of many of the models investigated that
included cohort effects; in particular, parameters in such models were
found to be sensitive to the fitting period. Furthermore,
\citet{Palin2016} has identified some concerns regarding potentially
spurious quadratic patterns in cohort effects in several of the models
discussed above, caused by variation in mortality improvement rates by
age being captured in the cohort effect.

\citet{rhglm} identify commonalities between the Lee-Carter model and
their Generalised Linear Model (GLM) approach to mortality modelling
focusing on mortality reduction factors. Instead of modelling declines
in mortality using a bi-linear term \(b_x\kappa_t\), however, Renshaw
and Haberman include a term \(b_xt\) that is linear in time, simplifying
the fitting process. The \(b_x\) parameters now represent age-specific
mortality improvements, where improvements are defined as differences in
log-mortality. In as similar vein, and building on the cohort
enhancement proposed by \citet{Renshaw2006}, an Age-Period-Cohort model
for Improvements (APCI) has been developed by the Continuous Mortality
Investigation (CMI) \citep{CMI2016}. However, this forces a
deterministic convergence to user-specified long-term rates of mortality
improvement rather than using time-series methods for forecasting.
\citet{Richards2017}, however, do provide full stochastic forecasts
using the APCI model by fitting time-series models to the period and
cohort effects, and also find that this model fits the data better
in-sample than either the APC or Lee-Carter models.

The smoothing of mortality rates is important in forecasting
applications to avoid roughness in the age profile of log-mortality due
to random variation being perpetuated into the future. A number of
smoothing models have thus been proposed. \citet{Hyndman2007} approach
the problem of mortality forecasting from within the functional data
paradigm. From a different perspective, \citet{Currie2004} fits a
two-dimensional P-spline to mortality, and produces forecasts by
extending the spline into the future. The penalisation of differences in
the basis function coefficients used in the P-spline method to ensure
smoothness in-sample also provides for extrapolation. Although this
model fits the data well, forecasts wholly dependent on extrapolation
from splines are likely to be over-sensitive to data and trends at the
forecast origin.

Bayesian methods are also increasingly being employed for mortality
forecasting in order to incorporate prior knowledge about underlying
processes, and provide distributions of future mortality risk accounting
for multiple sources of uncertainty. \citet{Girosi2008} demonstrate
methods for mortality forecasting within a Bayesian framework that allow
for smoothing the underlying data together with borrowing strength
across regions, as well as jointly forecasting cause-specific mortality.
\citet{Wisniowski2015} use the Lee-Carter method for all three
components of demographic change (fertility, mortality and migration),
again using Bayesian methods to obtain predictive probability
distributions.

The method developed in this paper combines elements of many of the
approaches above, including allowing for smooth functions of age and
cohort, while providing stable estimates of mortality at extreme ages
and avoiding some of the problems caused by lack of robustness in
parameter estimation discussed above. The model also shares some
features with the APCI model of \citet{Richards2017}, particularly in
the structure of the main part of the model. However, there are some
significant points of difference; the model described here applies to
the entire age range, and adopts a Bayesian approach to account for all
sources of uncertainty.

\subsection{Structure}\label{structure}

The remainder of the paper is structured as follows: the next section
(\ref{sec:model}) sets out the features of the model used in later
sections. Section \ref{sec:data_est} details the data used and the
estimation procedure. Section \ref{sec:fit} presents the posterior
distributions of the GAM components and provides predictive
distributions for log-rate forecasts, and Section \ref{sec:stacking}
displays posterior distributions combined over several alternative
models on the basis of in-sample predictive performance, using the
method of \citet{Yao2017}. Section \ref{sec:joint} presents an
alternative model where the sexes are fitted jointly, while Section
\ref{sec:holdback} compares out-of-sample performance of the single-sex
and joint models, using the years 2004-2013. Section
\ref{sec:ons_results} contrasts forecasts from the joint model with
those made by the UK Office for National Statistics
\citep{OfficeforNationalStatistics2016c}, and the final section offers
some conclusions and directions for future work.

\section{Model Description}\label{sec:model}

\subsection{Bayesian Generalised Additive
Models}\label{bayesian-generalised-additive-models}

Generalised Additive Models provide a flexible framework for modelling
outcomes where the functional form of the response to covariates is not
known with certainty, but is expected to vary smoothly. The general form
for such models is as follows \citep{Wood2006}:

\[
g(E(y_i)) = \mathbf{x}_i \boldsymbol{\theta}  +  s_1(x_{i1}) + s_2(x_{i2}) + \; ...
\]

Here, the expectation of the outcome \(y\), possibly transformed by link
function \(g(.)\), is modelled as the sum of a purely parametric part
\(\mathbf{X_i} \boldsymbol{\theta}\) and a number of smooth functions of
covariates \(s(.)\). A number of possible choices exist for the
implementation of the individual smooth functions, but P-splines are
chosen in this case. P-splines are appealing because they are defined in
terms of strictly local basis functions, with the domain of each
function defined by a set of knots spread across the covariate space
\citep{Wood2006}. Following the Bayesian P-splines approach of
\citet{Lang2001}, prior distributions are used to represent a belief
that adjacent P-spline covariates \(\boldsymbol{\beta}\) will be close
to one another. Multivariate normal prior distributions are used, with
the covariance matrix constructed from two matrices, \(A\) providing a
penalty on the first differences of the vector of coefficients
\(\boldsymbol{\beta}\), and \(B\) penalising the null-space of \(A\)
ensuring that the resulting prior is proper \citep{Wood2016}:

\begin{equation}
\begin{aligned}
s(x) &= \boldsymbol{\beta}^{T}\textbf{b}(x) \\
\boldsymbol{\beta} &\sim \text{MVN} \left(\textbf{0}, \; \left[\frac{1}{\sigma^{2}_A} A + \frac{1}{\sigma_B^{2}}B \right]^{-1} \right) .\\
\end{aligned}
\label{eq:coef_priors}\end{equation}

\subsection{Generalised Additive Models for Mortality
Forecasting}\label{generalised-additive-models-for-mortality-forecasting}

The method of mortality forecasting developed in this paper fits a GAM
to the majority of the age range, whilst applying separate parametric
models to older age groups and to infants. This allows a flexible but
smooth fit where the data allow, and imposes some structure on the model
where data are sparse, particularly at very high ages. Deaths \(d_{xt}\)
are considered to follow a negative binomial distribution parameterised
in terms of the mean, which in this case is equal to the product of the
relevant exposure \(E_{xt}\) and expected death rate \(m_{xt}\). The
dispersion \(\phi\) captures additional variance relative to the Poisson
distribution:

\[
\begin{aligned}
d_{xt} &\sim \text{Neg. Binomial}(E_{xt}m_{xt}, \; \phi) \\
p(d_{xt} | m_{xt}, E_{xt}, \phi) &= \frac{\Gamma(d_{xt} + \phi)}{d_{xt}! \; \Gamma(\phi)} \left(\frac{E_{xt}m_{xt}}{E_{xt}m_{xt} + \phi} \right)^{d_{xt}}  \left(\frac{\phi}{E_{xt}m_{xt} + \phi} \right)^{\phi}.
\end{aligned}
\]

An Age-Period-Cohort GAM for the log-mortality improvement ratios
\(\text{log}(\frac{m_{xt}}{m_{x(t-1)}})\) could be expressed with
P-spline based smooth functions for age and cohort improvements, and an
additional period component \(\kappa\):

\begin{equation}
\text{log}(\frac{m_{xt}}{m_{x(t-1)}}) = s_{\beta}(x) + s_{\gamma}^{*}(t-x) + \kappa^{*}_t
\label{eq:APC}\end{equation} .

An equivalent expression of this model can be made in terms of mortality
rates rather than mortality log-improvement ratios

\begin{equation}
\text{log}(m_{xt}) = s_{\alpha}(x) + s_{\beta}(x)t + s_{\gamma}(t-x) + \kappa_t,
\label{eq:log_m}\end{equation} with the cohort and period terms now
accumulated versions of their equivalents in the Equation \ref{eq:APC}.
This is the model used in the estimation process. There are now two
smooth functions of age: \(s_{\alpha}(x)\), which describes the
underlying shape of the log mortality curve; and \(s_{\beta}(x)\) which
describes the pattern of (linear) mortality improvements with age. Knots
are spaced on regular intervals in both the age and cohort direction
(every 4 years), with 3 knots placed outside the range of the data at
either end of the age range, allowing for proper definition of the
P-spline at the edge of the data.

In common with other models involving age, period, and cohort elements,
constraints are needed in order to identify the different effects
because of the linear relationship between the three components. To this
end, the cohort component \(s_{\gamma}(t-x)\) is constrained so that the
first and last components are equal to zero, and the sum of effects over
the whole range of cohorts is zero. The period components \(\kappa_t\)
are similarly constrained to sum to zero and to display zero growth over
the fitting period. The full set of constraints is thus:
\begin{equation}
\begin{aligned}
\sum^{T}_{t=1} \kappa_t &= 0 ; \; 
\sum^{T}_{t=1} t\kappa_t = 0 ; \\
\sum^{C}_{c=1} s_{\gamma}(c) &= 0 ;  \;
s_{\gamma}(1) = 0 ; \;
s_{\gamma}(\text{C}) = 0 \;,
\end{aligned}
\label{eq:constraints}\end{equation} with \(C\) here indicating the most
recent cohort and \(T\) the latest year. These constraints ensure that
linear improvements in mortality with time are estimated as part of the
\(s_{\beta}(x)\) term.

For older ages, a parametric model is adopted due to the sparsity of the
data in these regions -- the additional structure provided by specifying
a parametric form guards against over-fitting and instabilities in this
age range: \begin{equation}
\begin{aligned}
m_{xt} = \frac{\text{exp}(\beta^{old}_0 + \beta^{old}_1x + \beta^{old}_2t + \beta^{old}_3xt)}{1 + \text{exp}(\beta^{old}_0 - \text{log}(\psi) + \beta^{old}_1x + \beta^{old}_2t + \beta^{old}_3xt)} \; \text{exp}(s_{\gamma}(t-x) +\kappa_t)\\
\forall x:x \geq x_{old}.
\end{aligned}
\label{eq:old_age}\end{equation}

A logistic form is used, allowing mortality rates to tend toward a
constant \(\psi\) as age increases, as in the model in
\citet{Beard1963}. Such a pattern in mortality at the population level
has some theoretical justification, as it can result when heterogeneity
(`frailty') is applied to rates that follow a log-linear Gompertz
mortality model at the individual level, and this frailty is assumed to
be distributed amongst the population according to a gamma distribution
\citep{Vaupel1979}. In the life-table context, \citet{Dodd2018} found
that the logistic form performed better than the log-linear equivalent
when assessed using cross-validation techniques. Linear age and time
effects are included in the old-age model, together with an interaction
term, and the cohort and period effects are held in common with the
model applied to younger ages and are applied multiplicatively to the
logistic model.

Constraints are also applied to the parameters of the old age model to
ensure that the derivative of the parametric part of the model with
respect to age (ignoring the period and cohort effects) is never less
than zero; this reflects our prior belief that mortality should not
decrease with age after middle-age. The constraints required are as
follows, with \(H\) describing the most distant time for which forecasts
are desired:

\begin{equation}
\begin{aligned}
\beta^{old}_1 &> 0 \\
\beta^{old}_2 &< 0 \\
\beta^{old}_3 &> - \frac{\beta^{old}_1}{\text{H}}  \;.
\end{aligned}
\label{eq:old_constraints}\end{equation}

Infant mortality is also excluded from the GAM, as it behaves
differently from mortality at other ages. The model for infants is given
a similar structure to the old age model, except that the period effect
\(\kappa_t\) is excluded, as variation in infant mortality with time
does not appear to follow the same pattern as it does over the rest of
the age range.

\begin{equation}
\text{log}(m_{0t}) = \beta_1^{0} + \beta^{0}_1t +  s_{\gamma}(t) \;.
\label{eq:infant}\end{equation}

The period-specific effects \(\kappa_t\) in Equations \ref{eq:log_m} and
\ref{eq:old_age} are common across ages and capture deviations from the
linear trend described by the smooth improvements \(s_{\beta}\). These
effects are not modelled as smooth, as they may capture effects such as
weather conditions or infectious disease outbreaks that would not be
expected to vary smoothly from year to year. The innovations in these
period effects \(\epsilon\) are given a normal prior with variance
\(\sigma_{\kappa}\), so that

\begin{equation}
\begin{aligned}
\kappa_t  &= \kappa_{t-1} +  \epsilon_t  \\
\epsilon_{\kappa} &\sim \text{Normal}(0, \sigma_{\kappa}^{2}) \;.
\end{aligned}
\label{eq:kappa}\end{equation}

However, these effects are constrained in order to identify the APC
model, so we need to account for this by conditioning on the two period
constraints given in Equation \ref{eq:constraints}. This is achieved by
transforming the \(\pmb{\epsilon}\) parameters using a matrix \(Z\),
constructed so that the final \(T-2\) parameters remain unchanged, but
the first two transformed parameters will equal zero if the constraints
on the cumulative sum of the \(\pmb{\epsilon}\) series hold (see
Appendix). The resulting vector \(\eta\) has a multivariate normal
distribution

\begin{equation}
\begin{aligned}
\pmb{\eta} &= Z\pmb{\epsilon} \\
\pmb{\eta} &\sim \text{Multivarate Normal}(\pmb{0}, ZZ^{T}\sigma_{\kappa}^{2}).
\end{aligned}
\label{eq:eta}\end{equation}

A distribution conditioning on the first two elements of \(\pmb{\eta}\),
denoted \(\eta_{\dagger}\), equaling zero can be obtained using standard
results for the multivariate normal. This conditional prior on
\(\pmb{\eta^{*}}\) (which contains the last \(T-2\) elements of
\(\pmb{\eta}\)) is the distribution used for sampling, and the full set
of values of \(\pmb{\epsilon}\) can then be recovered deterministically

\begin{equation}
\begin{aligned}
\pmb{\eta} &= \begin{bmatrix}
\pmb{\eta^{\dagger}} \\
\pmb{\eta^{*}}
\end{bmatrix}\\
\pmb{\eta{*}} | (\pmb{\eta^{\dagger}} = \textbf{0}) &\sim \text{N} (0, \Sigma_{**} - \Sigma_{*\dagger}\Sigma_{\dagger\dagger}^{-1}\Sigma_{\dagger*}) \\
\Sigma &= ZZ^T \sigma_{\epsilon}^{2} \\
\pmb{\epsilon} &= Z^{-1} \begin{bmatrix}
\pmb{0} \\
\pmb{\eta^{*}}
\end{bmatrix},
\end{aligned}
\label{eq:conditional}\end{equation} where subscripts on the covariance
matrices indicate partitions so that \(\Sigma_{*\dagger}\) is the
sub-matrix of \(\Sigma\) with rows corresponding to \(\eta^{*}\) and
columns to \(\eta^{\dagger}\). For forecasts, innovations of the period
coefficients are unconstrained and so have independent normal
distributions with variance \(\sigma_{\kappa}^{2}\).

The same method is used to define a distribution for the innovations in
the basis functions coefficients for the cohort spline, accounting for
the cohort constraints in Equation \ref{eq:constraints}. In contrast to
the period effects, however, the transformation matrix used accounts for
the fact that the constraints apply to the resulting smooth function and
not the coefficient values themselves. Knots for the basis functions of
the cohort smooth are evenly spaced along the range of cohorts to be
estimated, so forecasts of future cohort values can be obtained by
drawing new coefficient innovations from the normal distribution with
mean zero and variance \(\sigma_{\gamma}^{2}\). Full details are given
in the appendix.

Priors for the model hyper-parameters are generally vague, although not
completely uninformative: \[
\begin{aligned}
\beta^{old} &\sim \text{Normal}(0, 100) \\
\beta^{0} &\sim \text{Normal}(0, 100) \\
\sigma_A &\sim \text{Normal}_+(0, 100) \\
\sigma_B &\sim \text{Normal}_+(0, 100) \\
\sigma_{\kappa} &\sim \text{Normal}_+(0, 100) \\
\sigma_{\gamma} &\sim \text{Normal}_+(0, 100) \\
\phi &\sim U(-\infty, \infty) \\
\psi &\sim \text{Log Normal} (0, 1). \\
\end{aligned}
\] The adoption of weakly informative priors aims to capture something
about the expected scale and location of the parameters in question;
this aids convergence of the Monte Carlo Markov Chain (MCMC) samples,
but with reasonable amounts of data should not affect the final
inference to any great extent \citep{Gelman2014a}. The scale of the data
and covariates is also important in determining the interpretation of
these priors; the use of standardised age and time indexes means that
regression coefficients are unlikely to take large values. The use of
addition symbol as a subscript appended to the normal distribution,
\(\text{Normal}_+\), indicates that only the positive part of the normal
distribution is used, therefore referring to a half-normal distribution.

\section{Estimation}\label{sec:data_est}

Samples from the posterior distributions of the parameters and rates
were drawn using Hamiltonian Monte Carlo (HMC) and specifically using
the \texttt{stan} software package \citep{StanDevelopmentTeam2015}. Stan
and its interface in the \texttt{R} programming language \citep{rman}
allows the construction of a HMC `No U-turns Sampler' (NUTS)
\citep{Hoffman2014} from a simple user specification of the Bayesian
model to be estimated. The code required to fit the model is provided in
the supplementary materials for this paper. HMC is a special case of the
more general Metropolis-Hastings algorithm for Markov Chain Monte-Carlo
sampling, and uses the derivatives of log-posterior with respect to the
parameters of interest in the sampling process, often allowing the
posterior to be traversed much more quickly than is the case under
standard methods \citep{Neal2010}. The model was fitted using Human
Mortality Database data for the UK from 1961--2013 (\citet{hmd}). The
first five cohorts (those born before 1856) are excluded, as exposures
are very low for these groups. Four parallel chains were constructed,
each with 8000 samples, and the first half of each chain was used as a
warm-up period (during which \texttt{stan} tunes the algorithm to best
reflect the characteristics of the posterior) and discarded. Parallel
chains were used to better assess convergence to the posterior
distribution; the diagnostic measure advocated by \citet{Gelman1992}
indicates that all parameters have converged to an acceptable degree.
The 16000 post-warm-up samples were `thinned' by a factor of 4 by
discarding three values in four to avoid excessive memory usage, leaving
4000 posterior samples for inference for each model.

\section{Initial Results}\label{sec:fit}

\begin{figure}[htbp]
\centering
\includegraphics{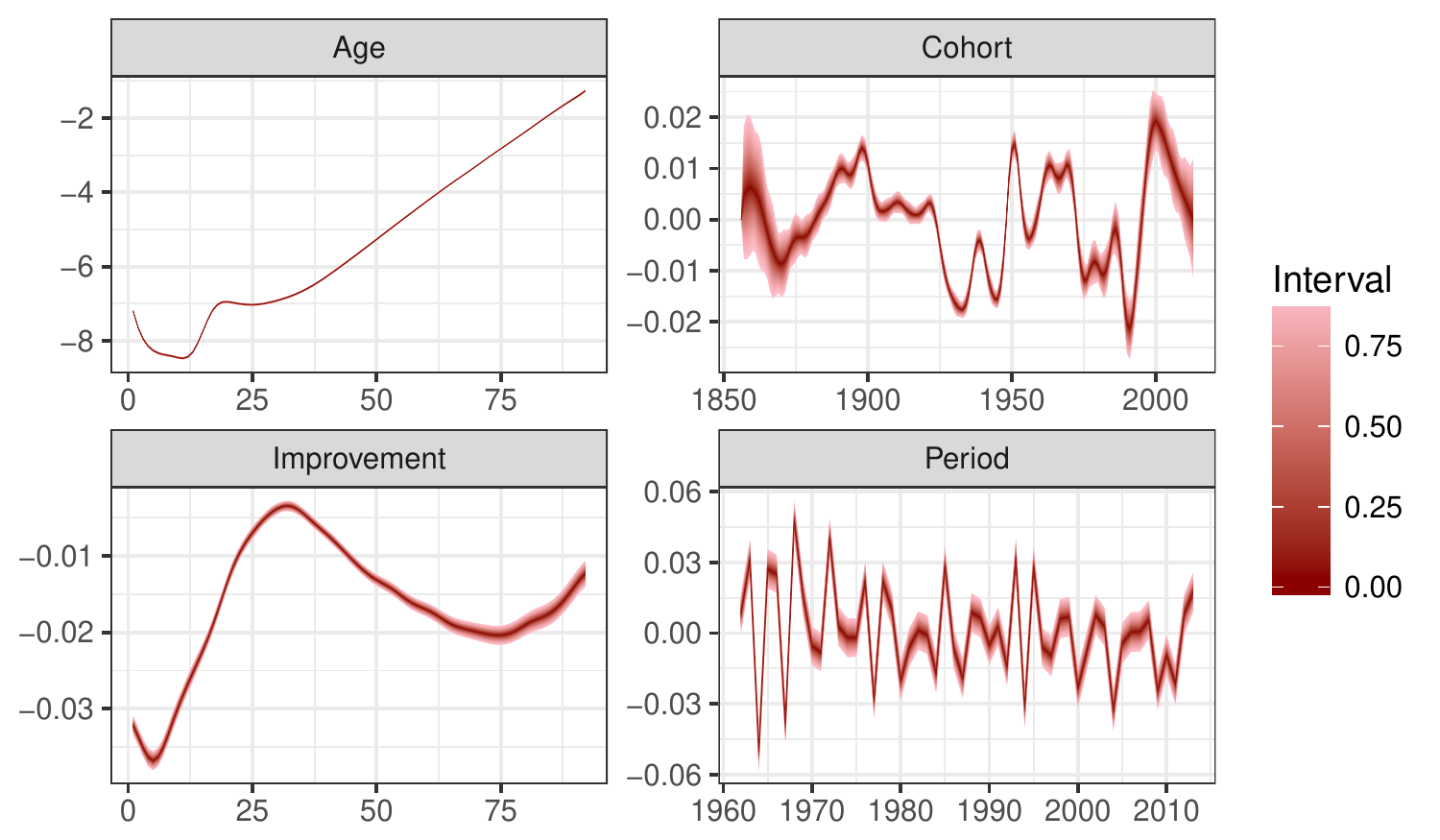}
\caption{GAM components, males, transition
point=93\label{fig:model_fit_m}}
\end{figure}

\begin{figure}[htbp]
\centering
\includegraphics{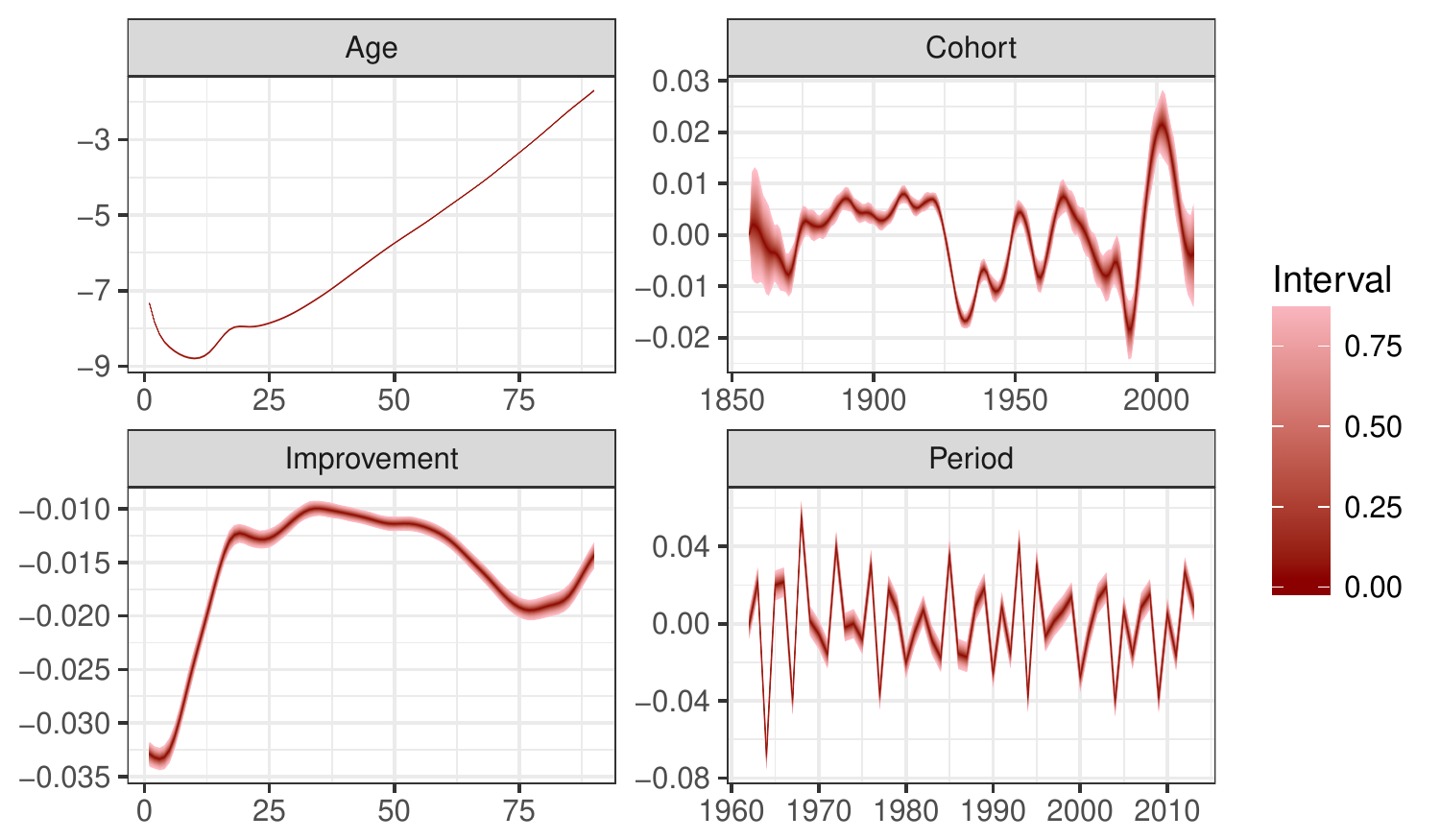}
\caption{GAM Components, females, transition
Point=91\label{fig:model_fit_f}}
\end{figure}

Some preliminary results are displayed in this section, conditional on a
particular choice for the point of transition between the GAM to the
parametric old-age model. Fitting a similar model to ONS data for
England and Wales for 2010-2012, \citet{Dodd2018} found using
cross-validation methods that the most probable points of transition
were age 91 for females and 93 for males. Samples were obtained for
models using these transition points, and the posterior distributions of
the parameters of the GAM model are given in Figures
\ref{fig:model_fit_m} and \ref{fig:model_fit_f} for males and females
respectively. The colour scheme in these plots identifies intervals
containing various proportions of the posterior density, so that the
deepest red represents the central 2\% interval, whilst 90\% of the
posterior density is contained between the lightest pink bands. The
distributions of mortality improvement rates for both males and females
display greater uncertainty at younger ages where there are fewer
deaths. As might be expected, uncertainty for cohort effects increases
for the oldest and most recent cohorts, as these have the fewest
data-points. Note that it is the differenced cohort and period effects
(\(s_{\gamma}^{*}(t-x)\) and \(\kappa^{*}_t\) from Equation
\ref{eq:APC}) that are plotted rather than their summed equivalents.

Differences between the sexes are most notable in the age-specific
component, for which the accident hump for young males is more
prominent, and in the improvement rates, for which males show lower
rates of improvement than females in their late 20s. Cohort and period
contributions to mortality decline show similar but not identical
patterns for each sex.

\begin{figure}[htbp]
\centering
\includegraphics{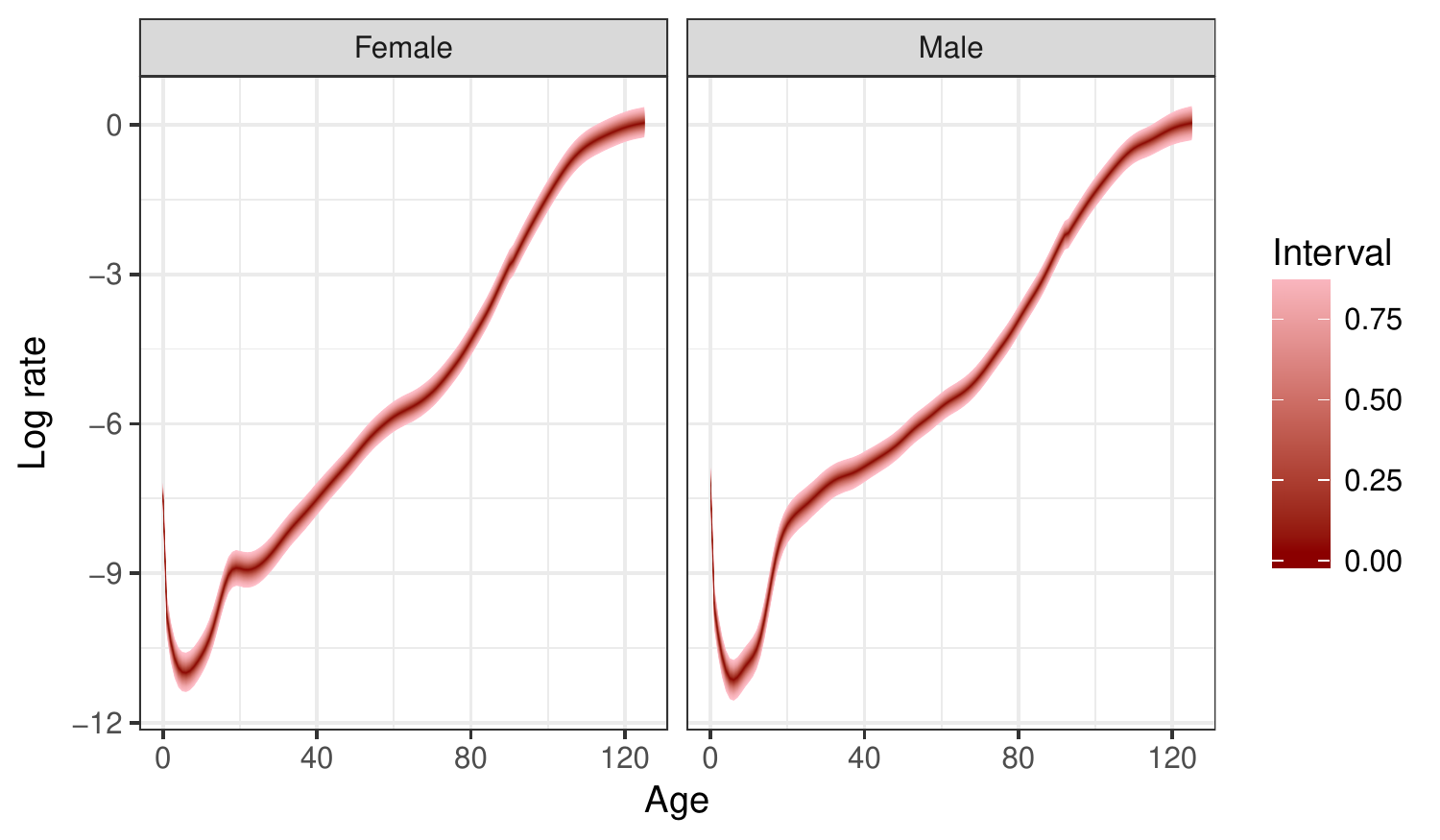}
\caption{Predictive distribution of log rates using single points of
transition, 2063\label{fig:fore_single}}
\end{figure}

Posterior distributions for log rates generated from this model fit the
data relatively closely. However, Figure \ref{fig:fore_single} displays
forecasts at fifty years into the future, which, while appearing
reasonable, contain a small discontinuity in the distribution of log
rates, particularly visible for males, at the point of transition
between the GAM and the parametric model. This suggests that some sort
of averaging over or combination of models using different transition
points might be advisable.

\section{Transition Points and Model Stacking}\label{sec:stacking}

The choice made regarding the age at which the model transitions from
the GAM (used over the majority of the age range) to the parametric
model for old ages is essentially arbitrary; we do not believe that
there is a switch between data-generating processes at some point
\(x_{old}\), but rather that the task of predicting mortality is better
served by two models. There is thus no `true' value for the point of
transition, and decisions regarding transition should be governed by
model performance. The methodology used in the latest English Life
Tables \citep{Dodd2018} used cross-validation to obtain posterior
weights over a set of models \(M\) defined by \(K\) different points of
transition, based on mortality data from 2010--2012. In that analysis,
age 91 for females and 93 for males are the most probable points of
transition, and the final predictive distribution was obtained by
averaging over models using the calculated weights. However, the model
described here differs to that used in \citet{Dodd2018} in that it
varies in time and applies to a period spanning many years, so the
question of the distribution of the transition between the parametric
model and the GAM must be revisited.

Separate models were therefore estimated for transition points ranging
from 80 to 95, and their accuracy was assessed using the
\emph{Leave-One-Out Information Criteria}, (LOOIC), developed by
\citet{Vehtari2015}. LOOIC is a measure of how well we might expect a
model to perform in predicting a data-point without including it in the
data used to fit the model. It is based on an approximation of the
Leave-One-Out (LOO) log point-wise predictive density
\(\sum^{n}_{i=1}\text{log}\;p(y_i | y_{-i})\), where the \(y_{-i}\)
subscript indicates a data-set excluding the \(i\)th observation,
\(\theta\) is a vector of parameters, and: \[
p(y_i | y_{-i}) = \int p(y_i|\theta)p(\theta|y_{-i})d\theta  \;.
\] Rather than fitting the model \(n\) times (once for every
data-point), \citet{Vehtari2015} provide a method for approximating the
LOOIC from just one set of posterior samples of the predictive density
computed from the full data-set, implemented within the \texttt{loo} R
package. This uses importance sampling to approximate the LOO
log-predictive density, correcting for instabilities caused by high or
infinite variance of the importance weights by fitting a Pareto
distribution to the upper tail of the raw weights.

The LOOIC scores for males and females for the models with transition
points \(k=[80,81,...,95]\) are given below in Figure \ref{fig:looic}.
Later cut-points tend to be preferred because the greater flexibility of
the GAM model gives higher LOOIC values even at relatively high ages,
although the absolute differences between the models are small. Models
with points of transition above age 95 are not considered, as this would
leave too few data-points with which to estimate the old-age model
effectively.

Although LOOIC is not a measure of \emph{forecast} performance as such,
as it is focused on how the model would perform at predicting
data-points contained within the original data-set and does not consider
the times at which data-points become available, it does provide an
indication of how well the specified models reflect the structure of the
data.

\begin{figure}[htbp]
\centering
\includegraphics{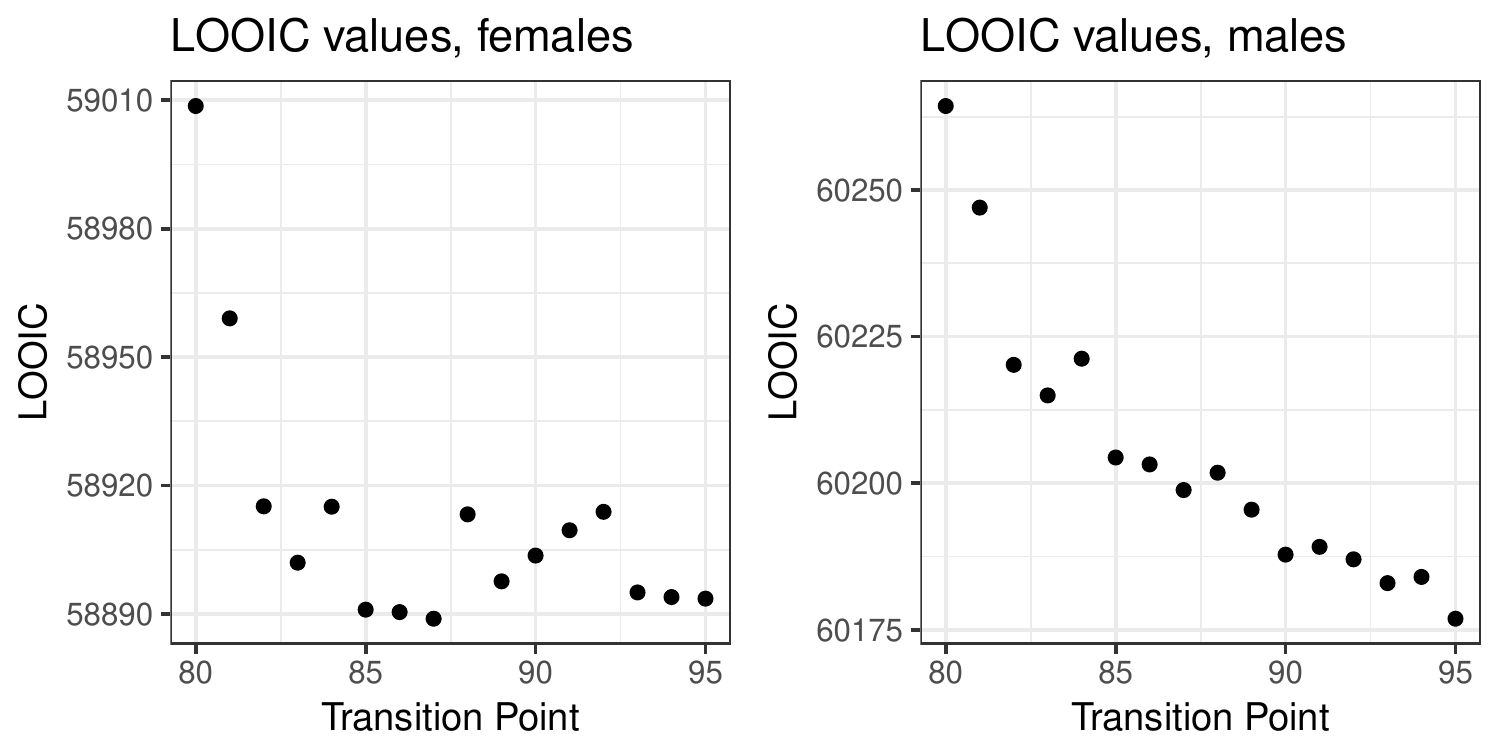}
\caption{LOOIC values for models using different transition
points\label{fig:looic}}
\end{figure}

\begin{figure}[htbp]
\centering
\includegraphics{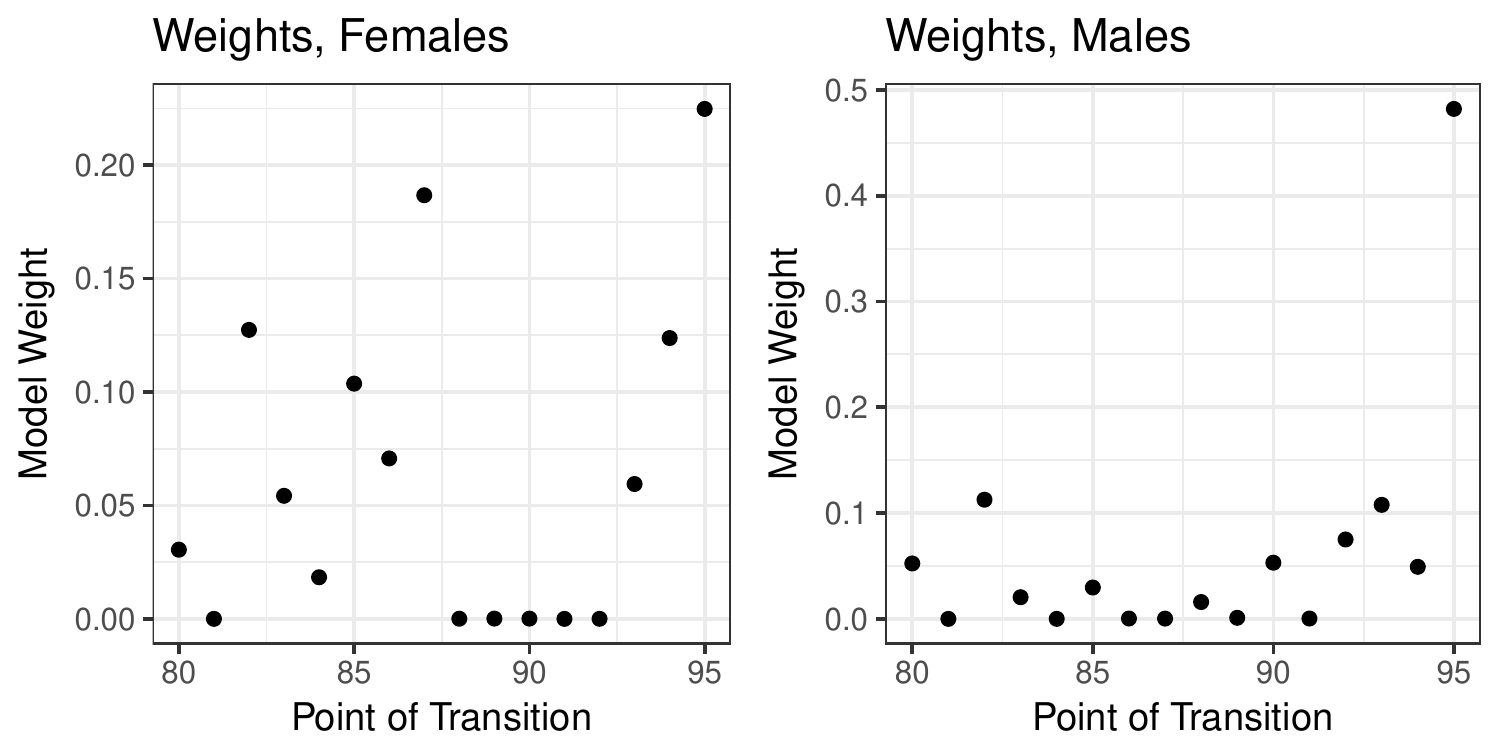}
\caption{LOOIC values for models using different points of
transition\label{fig:weights}}
\end{figure}

Following the work by \citet{Yao2017}, these LOOIC values can be used
for as the basis for `stacking' the predictive distributions of each
model to obtain a distribution which combines models in a principled
way, with weights determined by approximate cross-validation
performance. Stacking is often used for averaging over point estimates
in ensemble models, but \citet{Yao2017} extend the approach to apply to
combining distributions. More specifically, the weights \(\pmb{w}\),
elements of which corresponding to one of \(K\) possible models \(M_k\),
are estimated through the solution of the optimisation problem

\begin{equation}
\begin{aligned}
\underset{\pmb{w}}{\operatorname{argmax}} \sum_{i=1}^{n}\text{log}\left(\sum_{k=1}^{K} w_k \;p(y_i|y_{-i}, M_k)\right) \\ \text{s.t.} \; w_k > 0 ; \; \sum_{k=1}^{K}w_k=1 \; .
\end{aligned}
\label{eq:stacking}\end{equation} \citep[p.7]{Yao2017}, where
\(p(y_i | y_{-i}, M_k)\) is approximated using the LOOIC measure
described above. The form of the combined predictive distribution is
then \(\hat{p}(\tilde{y}|y) = \sum_{k=1}^{K}w_k p(\tilde{y}|y, M_k)\).
The estimated model weights are shown in Figure \ref{fig:weights}; the
greatest individual weight is given to models with the latest points of
transitions, reflecting the pattern in the LOOIC measure. Other models
with earlier transition points are also given weight, however,
reflecting that they perform well at predicting some data-points which
are not so well estimated by the late-transition model.

Samples from the combined posterior predictive distribution were
obtained using the estimated weights by sampling from the posterior
distribution associated with each model in proportion to its weight. The
resulting stacked forecasts are given below in Figure \ref{fig:stacked};
the discontinuities seen previously are now smoothed out through the
process of taking the weighted combination of distributions.

\begin{figure}[htbp]
\centering
\includegraphics{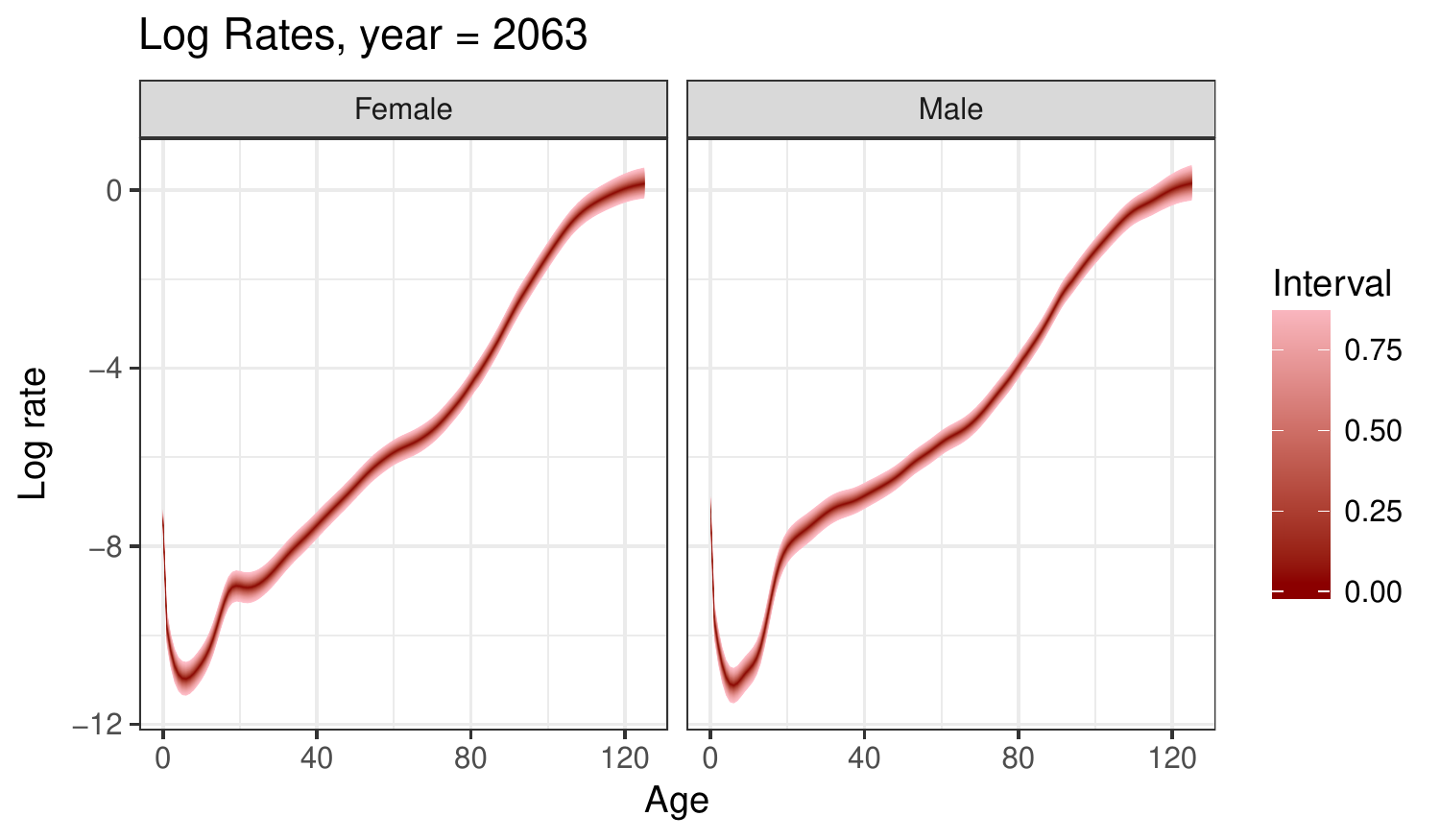}
\caption{Stacked forecasts for 2063, single-sex
models\label{fig:stacked}}
\end{figure}

\section{Jointly modelling male and female mortality}\label{sec:joint}

In the work described above, models for males and females are estimated
separately. However, much of what drives the underlying processes of
mortality and how it changes over time is likely to be common between
sexes. Thus, we may gain from borrowing strength across models and also
from explicitly representing covariances between parameters for each
sex, as in \citet{Wisniowski2015}. Because males tend to die sooner than
females, there are fewer data-points (that is, lower total exposure)
with which to estimate parameters in the old-age model. For this reason,
the parameter \(\psi\), representing the asymptote of the logistic
function in the old-age model, is now shared between sexes.

We also allow the innovations in the period effects \(\kappa_t\) to be
correlated, so that that joint forecasts can be generated accounting for
the fact that in potential futures where mortality for females is high,
it will tend to be high for males as well. The joint distribution for
the period innovations for both sexes, conditional on the constraints,
is obtained in a similar way as for the single-sex models, described in
Section \ref{sec:model}. Full details are given in the appendix.

\begin{figure}[htbp]
\centering
\includegraphics{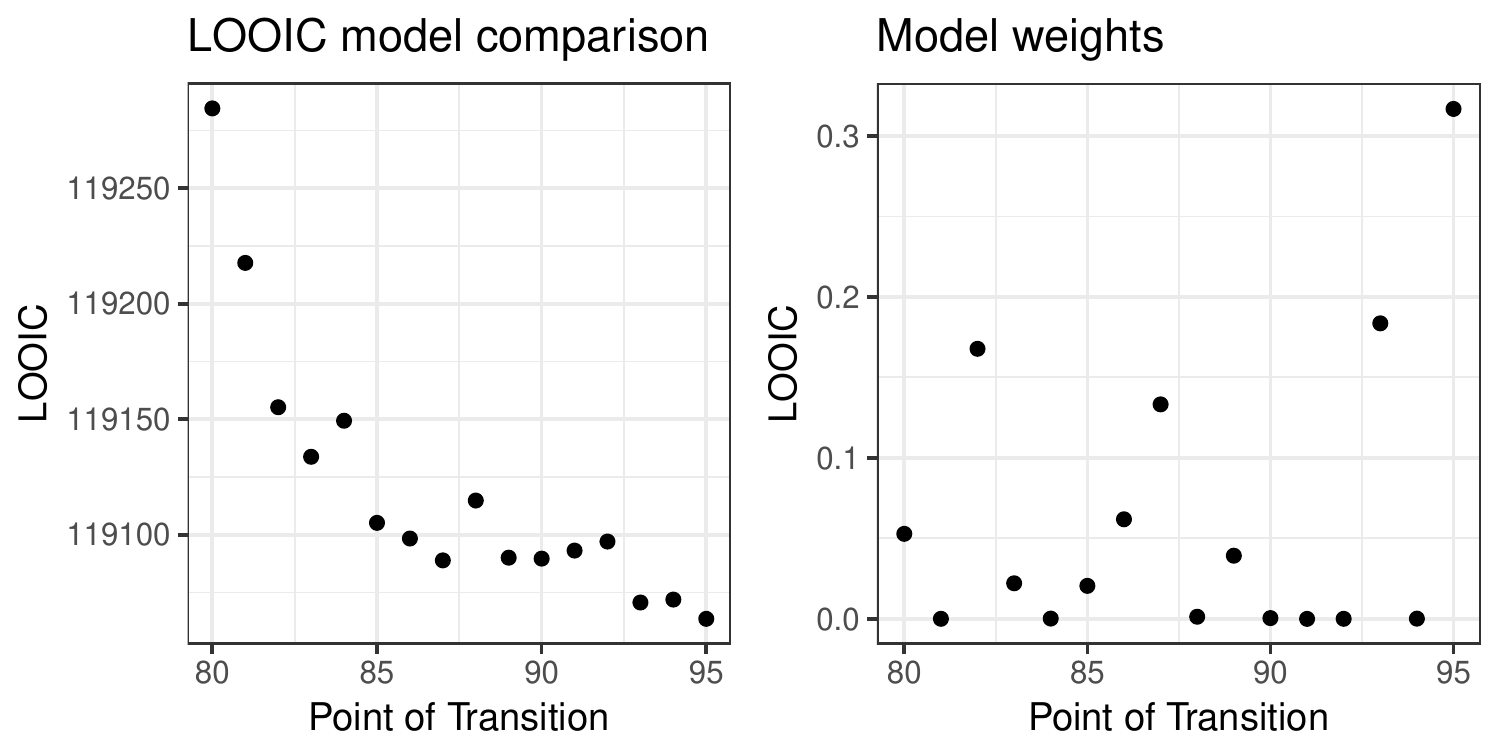}
\caption{LOOIC and model weights, joint-sex model\label{fig:joint_loo}}
\end{figure}

As before, LOOIC scores and model weights are obtained for the joint
model (Figure \ref{fig:joint_loo}). The pattern of LOOICs and weights
are similar to those for the separate models, with the highest
transition point obtaining most weight, but considerable weight also
attached to earlier transitions.

\begin{figure}[htbp]
\centering
\includegraphics{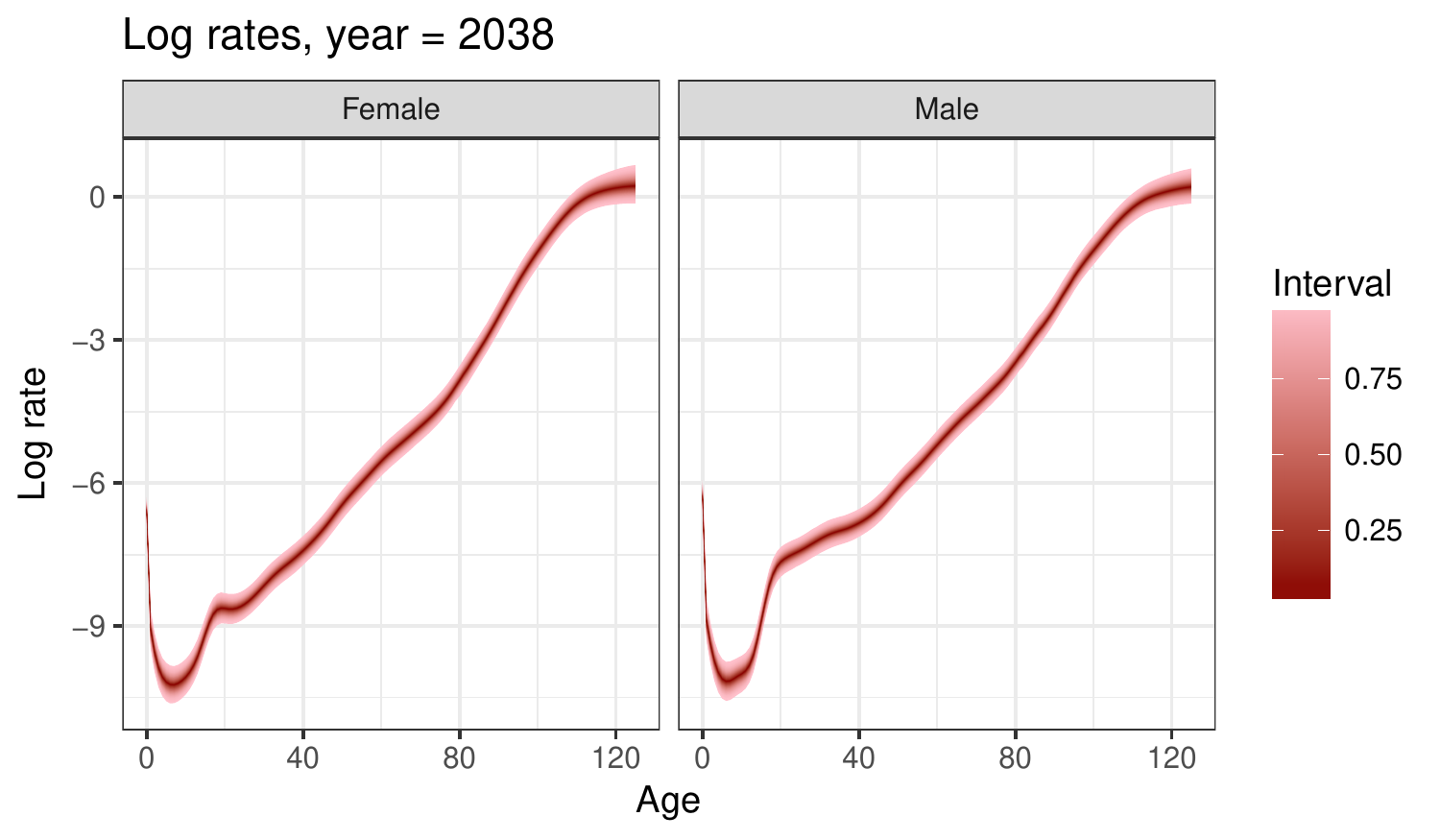}
\caption{Stacked forecasts from joint-sex model,
2038\label{fig:stacked_joint}}
\end{figure}

Joint forecasts of log-mortality are displayed in Figure
\ref{fig:stacked_joint}. The estimated correlation in the innovations of
the period effects (the off-diagonal elements of \(P\)) is high -
generally above 95\%.

\section{Model Assessment}\label{sec:holdback}

\begin{figure}[htbp]
\centering
\includegraphics{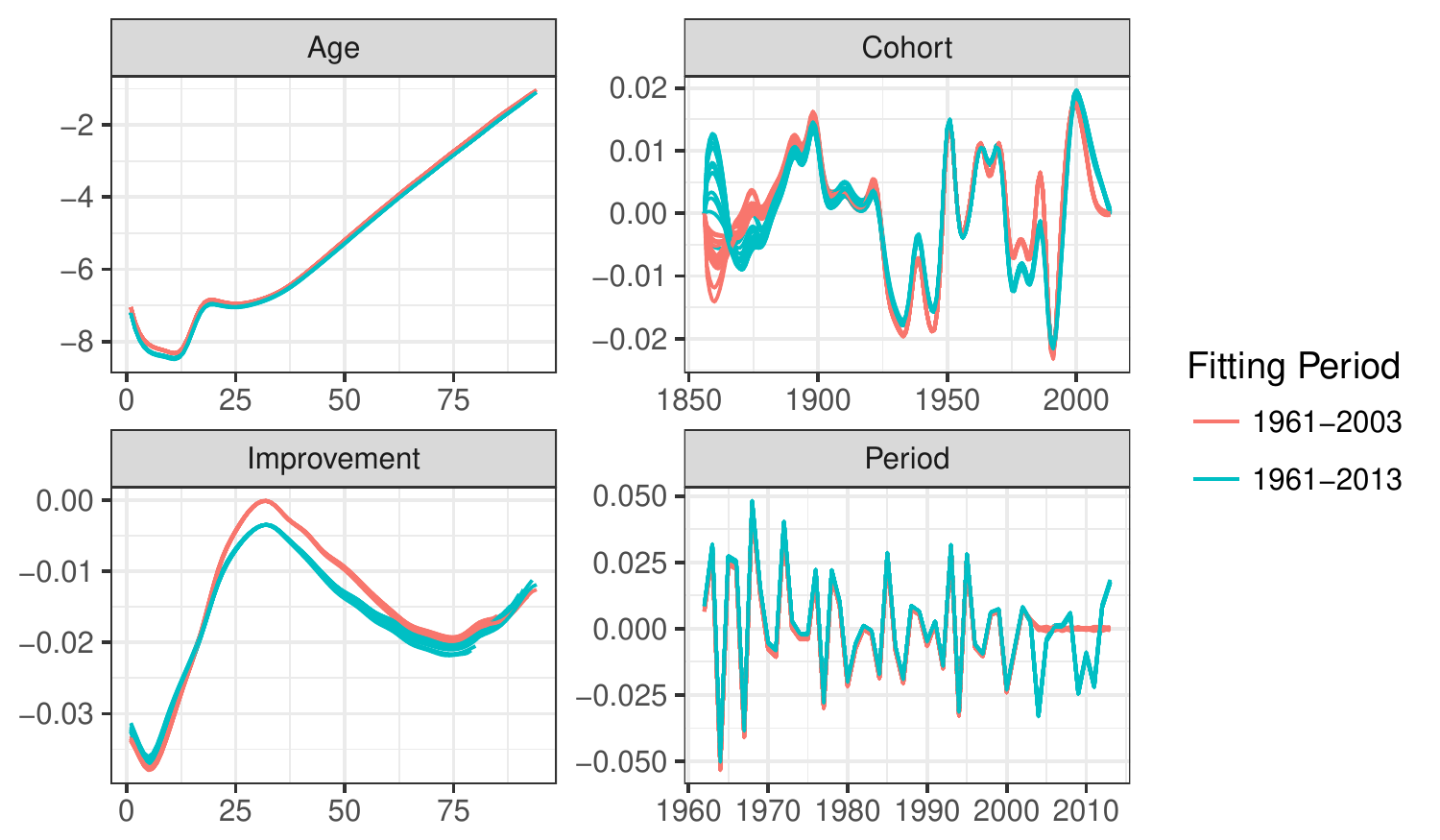}
\caption{Comparison of posterior means of GAM model components for
different fitting periods and transition points,
males\label{fig:parameter_comparison}}
\end{figure}

In order to assess the robustness and forecasting accuracy of the models
described above, fitting was conducted on a truncated data-set,
excluding the years 2004-2013. Robustness was then assessed by comparing
posterior means of the main smooth functions estimated on this reduced
data-set against the same quantities estimated on all the data. Figure
\ref{fig:parameter_comparison} displays such a comparison for males,
plotting posterior means for each point of transition and fitting
period. Estimates of period and cohort effects are relatively stable,
particularly in the interior of the data. While some differences are
evident in the pattern of improvements, the general shape of the curve
is notably similar, and the downward shift appears to reflect real
increases in the rate of mortality decline after 2003, particularly for
younger adults. The shape of the age effect is again very similar, and
the differing location of the smooth curve is accounted for by a change
in the location of the intercept of the time index in Equation
\ref{eq:log_m} for different data periods.

\begin{figure}[htbp]
\centering
\includegraphics{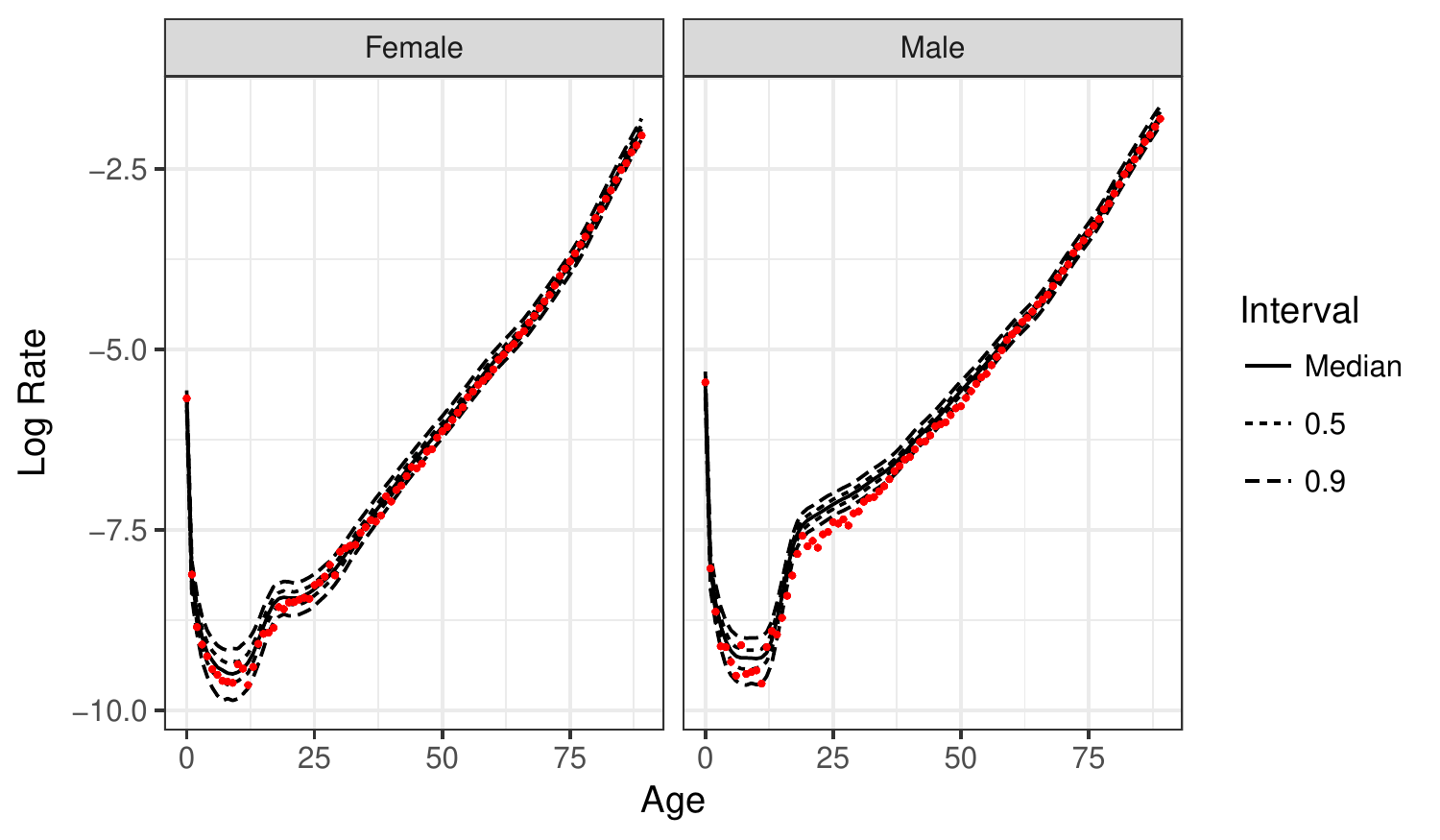}
\caption{Comparison of posterior predictive distributions for log-rates
against empirical observations, 2013, joint-sex
model.\label{fig:pred_ass}}
\end{figure}

\begin{figure}[htbp]
\centering
\includegraphics{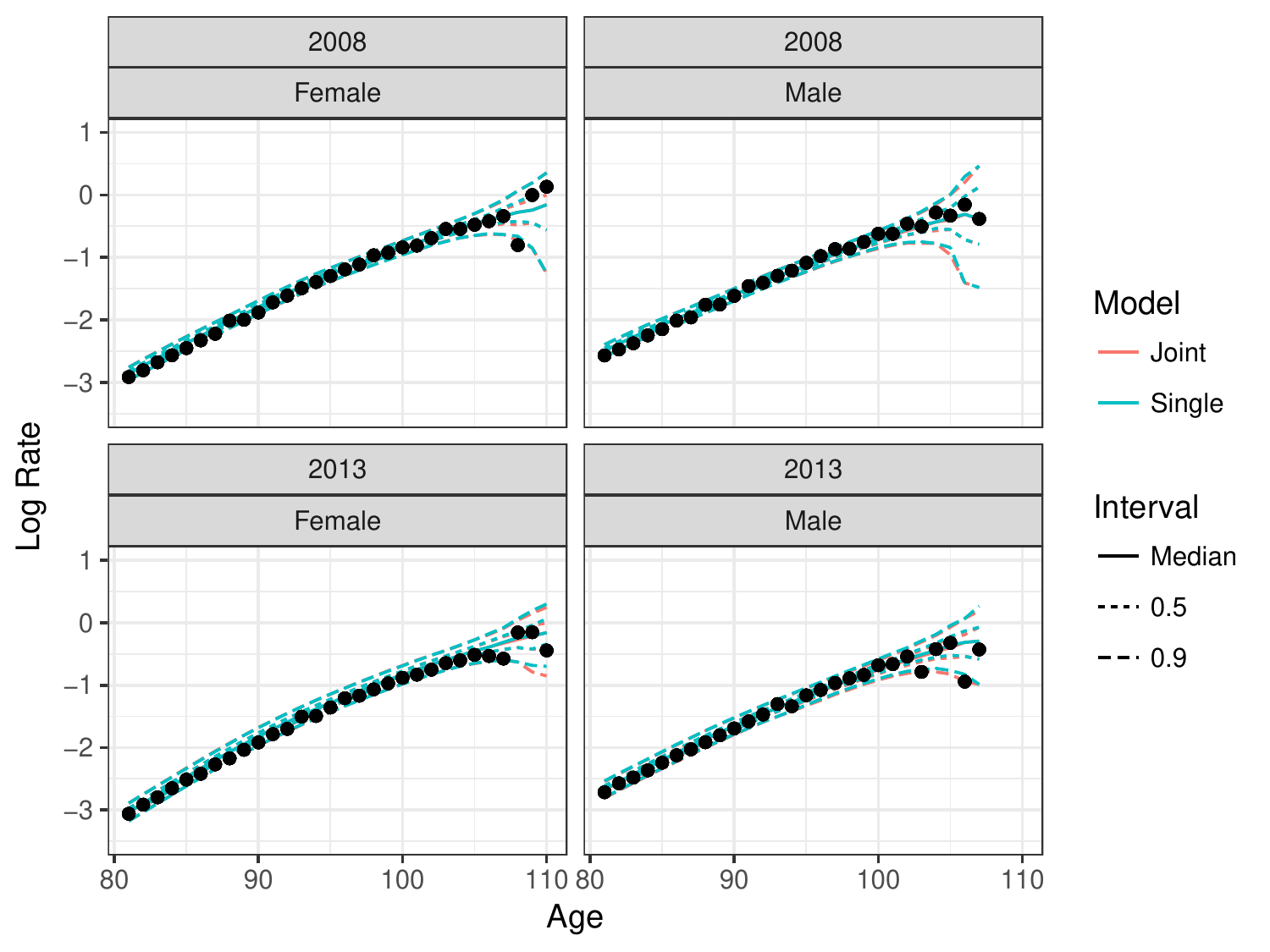}
\caption{Comparison of posterior predictive distributions for old-age
log-rates against empirical observations, 2008 and 2013, single and
joint-sex models.\label{fig:pred_ass_old}}
\end{figure}

Both the single and joint-sex models presented above appear to give
reasonable forecasts for future mortality. Figures \ref{fig:pred_ass}
and \ref{fig:pred_ass_old} display predictive distributions and
empirical rates for younger and older ages respectively. Comparing the
predictive posterior distributions against the observed outcomes, it is
evident that for most of the age range, empirical rates fall within the
90\% predictive interval. The exception is young adult males, between
the ages of about 15 and 40, for whom recent drops in mortality far
outpace those seen in the observed data 1961-2003. More formal
assessments of forecast performance are difficult, as we observe only
one correlated set of outcomes (that is, male and female log-rates
2004-2013).

Focusing on older ages (Figure \ref{fig:pred_ass_old}), we can see that
there are few differences between the predictive distributions of the
joint- and single-sex models, and those that are evident occur only at
high ages. In part, this may be because the weighting procedure works to
select models with similar properties. Other considerations may be taken
into account when deciding between the two models; the joint model is
more parsimonious in that fewer parameters are required to fit it, and
it allows for correlations in the paths of mortality by sex to be taken
into account. In contrast, the single sex model is less computationally
demanding, particularly with respect to memory, as each sex is fitted
and processed separately.

\section{Comparison to offical projections and
variants}\label{sec:ons_results}

The final stacked forecast from the joint model in the previous section
are now compared with forecasts produced by the United Kingdom Office
for National Statistics (ONS) in the 2014-based National Population
Projections (NPP) \citep{OfficeforNationalStatistics2016c}. These work
with the predicted probabilities of deaths \(q_x\) rather than the
central mortality rates \(m_x\); the former represents the probability
of dying by age \(x+1\) given that an individual attains age \(x\).
Posterior predictive samples of \(q_{xt}\) were acquired using the
approximation \begin{equation}
q_{xt} \approx 1-\text{exp}(-m_{xt}).
\label{eq:qx}\end{equation} As well as the principal ONS projection from
the 2014-based NPP, the variant projections involving high and low
mortality scenarios have been included, allowing some understanding of
how the existing indications of uncertainty resulting from different
projection assumptions compare with the fully Bayesian probability
distributions.

\begin{figure}[htbp]
\centering
\includegraphics{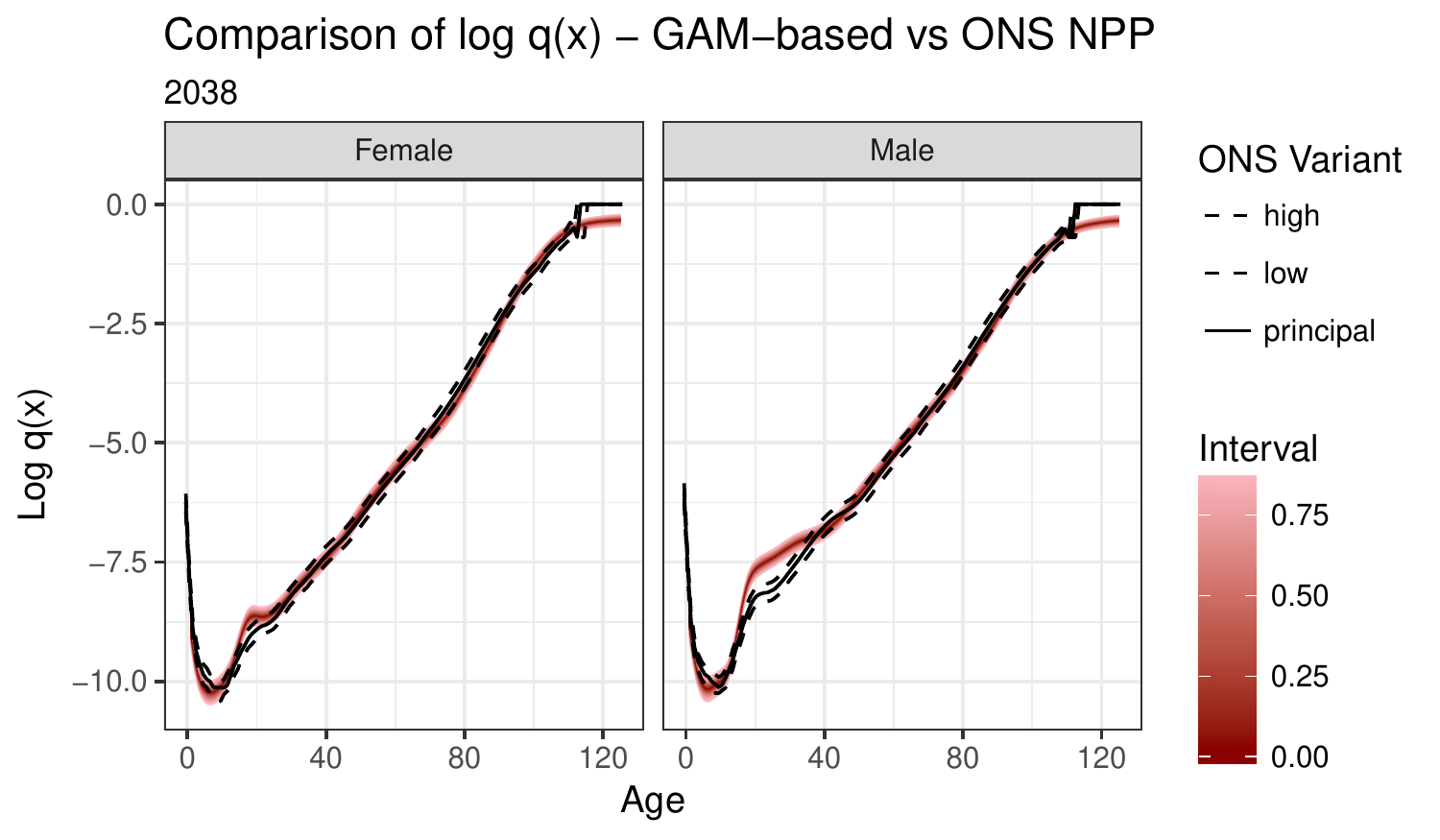}
\caption{Forecast log-probabilities of death for
2038\label{fig:forecast_25}}
\end{figure}

Figure \ref{fig:forecast_25} shows posterior distributions of
log-transformed death probabilities \(q_x\) for a forecast horizon of 25
years for both males and females, together with the equivalent
\(q_{x+0.5}\) quantities for the same year (2038) obtained from the ONS
2014-based NPP. For most of the age range, the forecasts are similar,
with the principal projection falling close to the median prediction
under the GAM-based model. However, the ONS model projects lower
mortality for young adults for both sexes, to the extent that the
principal projections fall outside the outermost 90\% predictive
interval of the probabilistic projections. This is due to a greater
weight given by the ONS methodology to more recent high improvement
rates at these ages \citep[see][ for more details regarding the ONS
methodology]{OfficeforNationalStatistics2016c}.

\subsection{Life Expectancy}\label{life-expectancy}

\begin{figure}[htbp]
\centering
\includegraphics{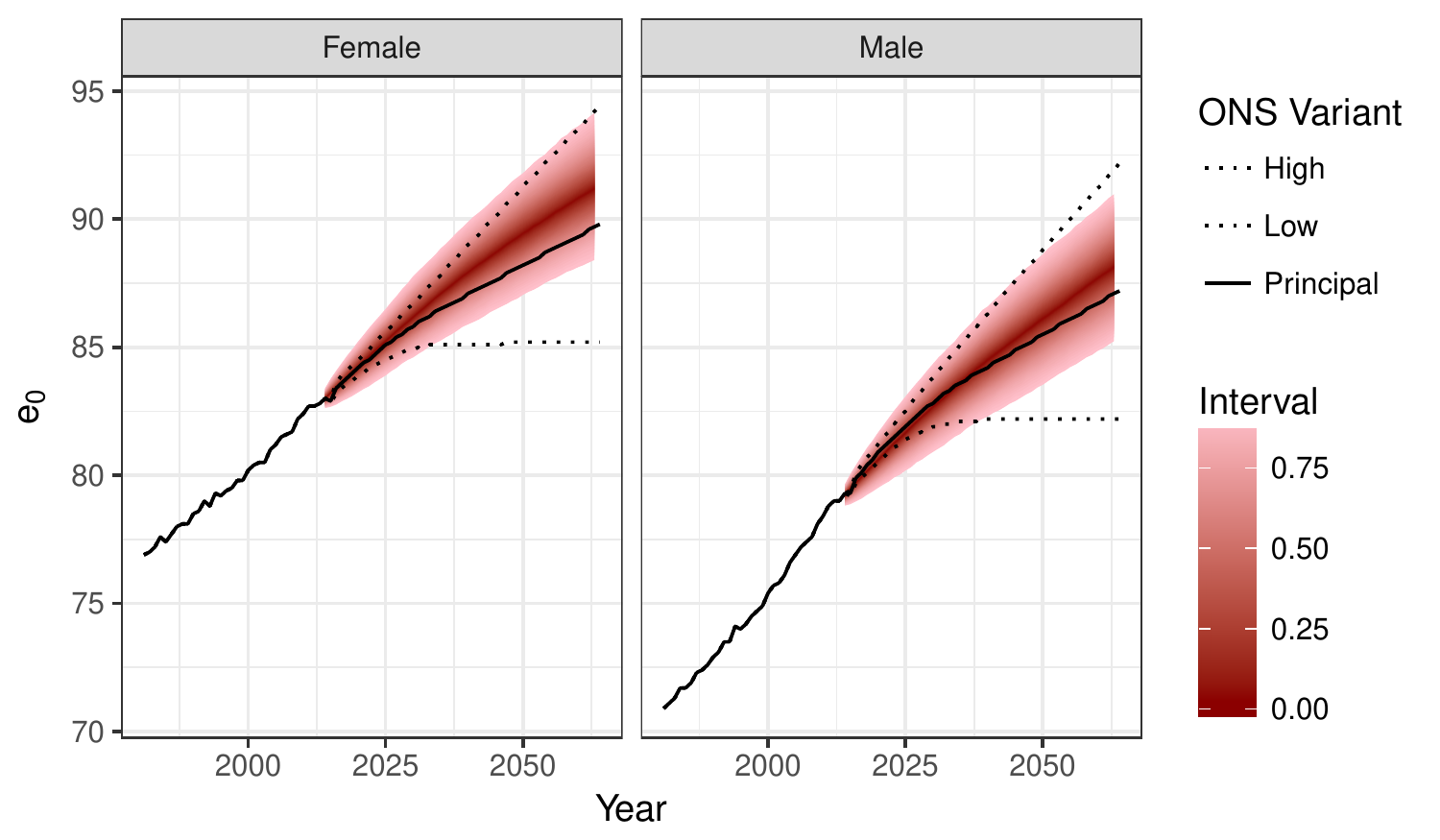}
\caption{Forecast Life Expectancy at Birth, ONS NPP and GAM-based
forecast\label{fig:e0_nu}}
\end{figure}

Period life expectancy at birth is a useful summary measure of the
mortality conditions in a given year. It captures the expected number of
years lived of a hypothetical individual who experiences a given
period's schedule of mortality rates over the course of their whole
life. Figure \ref{fig:e0_nu} compares the posterior distribution of life
expectancy at birth (\(e_{0}\)) from the jointly fitted GAM-based model
with the equivalent quantity from the NPP. The GAM-based forecasts
appear more optimistic than the ONS equivalent, with median life
expectancy higher than the principal ONS projection due to the lower
predictions of mortality at ages 70-95 under the GAM-based model. Figure
\ref{fig:e0_nu} also reveals that uncertainty in \(e_{0}\) initially
grows more quickly in the Bayesian approach developed above, in that the
gap between the high and low variants is much narrower than the fan
intervals for at least the first decade of the forecast. After 30 years,
however, the range spanned by the ONS variants becomes wider than 90\%
probabilistic interval from the GAM-based model. The uncertainty in the
probabilistic forecast reflects past variability in the observed data,
and from the comparison with hold-back data given in Figures
\ref{fig:pred_ass} and \ref{fig:pred_ass_old}, the calibration of this
uncertainty appear reasonable. As a result, we believe that the
probabilistic intervals provide a better indication of the uncertainty
around future life expectancy than the scenario-based equivalents, at
least in the short term, particularly as they have a readily
understandable interpretation in terms of probability.

\section{Discussion and Conclusion}\label{discussion-and-conclusion}

This paper details methodology for the fully probabilistic forecasting
of mortality rates, accounting for uncertainty in parameter estimates as
well as in forecasting. The approach uses a GAM to produce smooth rate
estimates at younger ages, and combines this with a parametric model at
higher ages where the data are more sparse, allowing rate estimates to
be obtained for extreme old ages. The use of Hamiltonian Monte Carlo
sampling and the \texttt{stan} software package allowed posterior
sampling to be conducted with reasonable efficiency.

Stacking predictive distributions following the approach of
\citet{Yao2017} provides a principled approach to avoiding a single
choice of transition point between these two sub-models governing
younger and older age ranges. These weights are based on approximate
Leave-One-Out cross-validation performance, and thus weight models based
on their ability to predict data contained in the original fitting
period. An alternative approach may be to fit models on a subset of
data, and produce weights based on model performance in forecasting data
at the end of the time period. However, this would involve additional
model refitting, and it may also be the case that such assessments are
overly sensitive to characteristics of the held-out data. Furthermore,
log-scores based on a single set of observed outcomes are likely to be
highly correlated, and thus rolling n-step-ahead forecasts may be
required to assess forecast performance robustly, which would
necessitate repeated model fitting with even greater computational
expense.

A comparison with ONS forecasts provides an indication of how Bayesian
predictive intervals compare with the deterministic scenario-based
indicators of forecast variability produced by ONS. For life expectancy
in particular, the probabilistic intervals are considerably wider over a
short time horizon than those suggested by the high and low mortality
scenarios. Future work could investigate the inclusion of expert opinion
in probabilistic mortality forecasting models like the one presented in
this paper. The NPP uses experts to provide target rates of mortality
improvement over longer time horizons (25 years)
\citep{OfficeforNationalStatistics2016c}, reflecting the fact that
extrapolative methods may prove inferior to expertise at this distance
into the future. A similar approach within a Bayesian framework would
have to consider that using expert opinion about future rates is
different from the standard approach of eliciting information about
model parameters directly. Work in \citet{Dodd2018b} describes one way
in which this could be achieved. Beyond this, there are also
opportunities to investigate the possibility of extending similar
methods to other demographic components, particularly fertility.

\bibliography{GAM}
\bibliographystyle{kluwer}

\end{document}

\maketitle

\emph{Supplemental Material to ``Projecting UK Mortality using Bayesian
Generalised Additive Models''}

\subsection*{R Code}\label{r-code}
\addcontentsline{toc}{subsection}{R Code}

Code to run all of the anaylsis in the paper is provided as a github
repository at \url{https://github.com/jasonhilton/mortality_bgam}.

\subsection*{Period Parameters}\label{period-parameters}
\addcontentsline{toc}{subsection}{Period Parameters}

This section describes how prior distributions for the period
constraints are constructed. Some details are repeated from the body of
the text for completeness. Period innovations are normally distributed
so that:

\[
\begin{aligned}
\kappa_t &= \kappa_{t-1} + \epsilon_t \\
\epsilon_t &\sim \text{Normal}(0, \sigma_{\epsilon}^{2}) \\
\end{aligned}
\] Defining \(T\) as the number of periods in the dataset, the \(2\) by
\(T\) matrix \(C\) describes the two constraints on the vector of
\(\pmb{\kappa}\) parameters, that they must sum to zero and show no
linear growth. \[
C = \begin{bmatrix}
1 & 1 & 1 & 1 & \dots & 1 & 1 \\
0 & 1 & 2 & 3 & \dots & T-2 & T-1 \\
\end{bmatrix}
\] When the constraints hold, \[
\begin{aligned}
C\pmb{\kappa} &= \pmb{0} \\
CS\pmb{\epsilon} &= \pmb{0},
\end{aligned}
\] where \(S\) is the cumulative sum matrix. A distribution for period
innovations conditional on the constraints is obtained by first
transforming \(\pmb{\epsilon}\) into a new set of parameters
\(\pmb{\eta}\), where the first two elements of \(\pmb{\eta}\) are zero
when the constraints hold, and the remaining elements are identical to
the equivalents in \(\pmb{\epsilon}\). The matrix \(Z\) used for this
transformation is a \(T\) by \(T\) identity matrix with the first two
rows replaced by the matrix \(CS\).

\[
\begin{aligned}
\pmb{\eta} &= \begin{bmatrix}
\pmb{\eta^{\dagger}} \\
\pmb{\eta^{*}}
\end{bmatrix} = Z\pmb{\epsilon} \\ 
\pmb{\eta} &\sim \text{MVN}(\pmb{0}, ZZ^{T}\sigma_{\kappa}).
\end{aligned}
\] By conditioning on the first two values of \(\pmb{\eta}\) (denoted
\(\pmb{\eta}^{\dagger}\)) equalling zero, we can find the distribution
of the last \(t-2\) elements of \(\pmb{\eta}\) using the standard
conditional relationship for multivariate normal variables. Conditioning
on these elements of \(\pmb{\eta}\) being equal to zero is equivalent to
conditioning on the constraints holding. We can therefore calculate the
values of the first two values of \(\pmb{\epsilon}\) by multiplying
\(\pmb{\eta}\) by the inverse of the \(Z\) matrix.

\[
\begin{aligned}
\Sigma &= ZZ^T{}\sigma^{2}_{\epsilon} \\
\pmb{\eta{*}} | (\pmb{\eta{\dagger}} = \textbf{0}) &\sim \text{MVN} (0, \Sigma_{**} - \Sigma_{*\dagger}\Sigma_{\dagger\dagger}^{-1}\Sigma_{\dagger*}) \\
\pmb{\epsilon} &= Z^{-1} \begin{bmatrix}
\pmb{0} \\
\pmb{\eta^{*}}
\end{bmatrix},
\end{aligned}
\] where subscripts on the covariance matrices indicate partitions so
that \(\Sigma_{*\dagger}\) is the sub-matrix of \(\Sigma\) with rows
corresponding to \(\pmb{\eta^{*}}\) and columns to
\(\pmb{\eta^{\dagger}}\).

\subsection*{Correlated period parameters for males and
females}\label{correlated-period-parameters-for-males-and-females}
\addcontentsline{toc}{subsection}{Correlated period parameters for males
and females}

When both sexes are modelled together, the period innovations for each
sex retain separate variance parameters, but are assumed to be joint
multivariate normal, with the correlation determined by parameter
\(\rho\). The joint distribution for the period innovations
\(\pmb{\epsilon}\) is therefore as follows:

\[
\begin{aligned}
 \begin{bmatrix}
           \pmb{\epsilon_{\kappa f}} \\
           \pmb{\epsilon_{\kappa m}}
         \end{bmatrix}
         &\sim \text{Multivariate Normal} \left( \textbf{0}, P \right) \\
         P &= \begin{bmatrix}
I_T \sigma^2_{\kappa f}& I_T\sigma_{\kappa m}\sigma_{\kappa f}\rho \\
I_T\sigma_{\kappa m}\sigma_{\kappa f}\rho&I_T \sigma^2_{\kappa m}
\end{bmatrix} \\
                  \rho &\sim \text{Beta}(1,1),
\end{aligned}\] where \(I_T\) is the identity matrix of dimension \(T\),
the number of periods, and \(\pmb{\epsilon_{\kappa m}}\) and
\(\pmb{\epsilon_{\kappa f}}\) refer to period innovations for males and
females respectively. The prior distribution of the period parameters
conditional on the constraints is obtained in a similar way to the
single-sex case. The individual innovation vectors for each sex are
first transformed to vectors \(\pmb{\eta_f}\) and \(\pmb{\eta_m}\). The
implied joint multivariate distribution can then be conditioned on the
constraints holding true for both sexes in the same way as before: \[
\begin{aligned}
\pmb{\eta} &= \begin{bmatrix}
\pmb{\eta^{f\dagger}} \\
\pmb{\eta^{f*}} \\
\pmb{\eta^{m\dagger}} \\
\pmb{\eta^{m*}}
\end{bmatrix}
 = X \pmb{\epsilon} \\
 X &= \begin{bmatrix}
Z & 0 \\
0 & Z 
\end{bmatrix} \\
\Xi &= XPX^{T} \\
\pmb{\eta} &\sim \text{MVN} (\pmb{0}, \Xi) \\
\pmb{\eta_{.*}} | (\pmb{\eta_{.\dagger}} = \textbf{0}) &\sim \text{N} (0, \Xi_{**} - \Xi_{*\dagger}\Xi_{\dagger\dagger}^{-1}\Xi_{\dagger*}). \\
\end{aligned}
\]

\subsection*{Cohort Parameters}\label{cohort-parameters}
\addcontentsline{toc}{subsection}{Cohort Parameters}

Cohort effects are modelled as p-splines, with innovations normally
distributed, so that \[
\begin{aligned}
s_{\gamma}(t-x) &= \pmb{\beta}^{\gamma}\pmb{b}(t-x) \\
\beta^{\gamma}_{i} &= \beta^{\gamma}_{i - 1} + \epsilon^{\gamma}_{i} \\
\epsilon^{\gamma}_{i} &\sim \text{Normal}(0, \sigma_{\gamma}^{2})
\end{aligned}
\] with \(B(.)\) giving the B-spline basis function, and \(i\) indexes
the individual basis functions. As with the period effects, a matrix
\(C\) is describes the constraints on the smooth function
\(s_{\gamma}\), applying to the first, second, and final elements of the
parameter vector. \[
\begin{aligned}
C = \begin{bmatrix}
1 & 0 & 0 & 0 & \dots & 0 & 0 \\
1 & 1 & 1 & 1 & \dots & 1 & 1 \\
0 & 0 & 0 & 0 & \dots & 0 & 1 \\
\end{bmatrix},
\end{aligned}
\] The constraints hold when \[
\begin{aligned}
C\pmb{s_{\gamma}} &= \pmb{0} \\
CBS\pmb{\epsilon^{\gamma}} &= \pmb{0}.
\end{aligned}
\] A new parameter vector \(\pmb{\eta}^{\gamma}\) is obtained as for the
period effects \[
\begin{aligned}
\pmb{\eta}^{\gamma} &= W\pmb{\epsilon}^{\gamma} \\
\pmb{\eta}^{\gamma} &= \begin{bmatrix}
\pmb{\eta^{\gamma\dagger}} \\
\pmb{\eta^{\gamma*}}\\
\pmb{\eta^{\gamma\ddagger}} \\
\end{bmatrix}\\
\pmb{\eta^{\gamma}} &\sim \text{MVN}(\pmb{0}, WW^{T}\sigma_{\gamma}^{2}).
\end{aligned}
\] The matrix \(W\) is an identity matrix with the first, second and
final rows replaced by the rows of the matrix \(CBS\), where \(B\) is
the matrix of basis functions \(\pmb{b}(.)\) evaluated for each cohort.
A distribution for \(\eta^{\gamma*}\) given the constraints can now be
constructed by conditioning on
\(\pmb{\eta}^{\gamma\dagger} = \pmb{\eta}^{\gamma\ddagger} = 0\), in the
same manner as for the period effects. New cohort basis function
innovations can be drawn from the normal distribution with mean \(0\)
and variance \(\sigma_{\gamma}^{2}\).

%\bibliography{GAM}
%\bibliographystyle{rss}